# Investigation of resistive switching in Au/$MoS_2$/Au using Reactive Molecular Dynamics and *ab-initio* quantum transport calculations

Ashutosh Krishna Amaram, Saurabh Kharwar, and Tarun Kumar Agarwal

*Abstract*— **In this work, we investigate the underlying physical mechanism for electric-field induced resistive switching in Au/$MoS_2$/Au based memristive devices by combining reactive Molecular Dynamics (MD) and *ab-initio* quantum transport calculations. Using MD with Au/Mo/S ReaxFF potential, we observe the formation of realistic conductive filament consisting of gold atoms through monolayer $MoS_2$ layer when sufficient electric field is applied. We furthermore instigate the rupture of the gold atom filament when a sufficiently large electric field is applied in the opposite direction. To calculate the conductance of the obtained structures and identify the High Resistance (HR) and Low Resistance (LR) states, we employ the *ab-initio* electron transport calculations by importing the atomic structures from MD calculations. For single-defect $MoS_2$ memristors, the obtained LRS, HRS current densities are in order of $10^7$ A/cm$^2$ which agrees reasonably well with the reported experiments.**

## I. Introduction

In recent years, Resistive Random Access Memory (RRAM) devices have gained significant attention for its application in analog in-memory computing [1]. Ti/$HfO_2$/TiN stacks have been successfully integrated in the CMOS chips [2]. However, the device-to-device and cycle-to-cycle variability remains a key problem in Transition Metal Oxides (TMO: $HfO_2$/$TaO_2$) based RRAM devices. Recently, atomically thin Two-Dimensional (2D) materials (e.g. $MoS_2$) based memristors have shown very low leakage currents compared to the traditional oxide based memristors along with large $R_{on}/R_{off}$ ratios and controlled variability [3,10]. The underlying switching mechanism behind the resistive switching in Au/$MoS_2$/Au memristor devices remains elusive. In the recent past, there has been significant effort to understand resistive switching with the aid of different theoretical studies. Li et al [4] studied the resistive switching in Au/$MoS_2$/Au memristors with the help of Density Functional Theory (DFT), where they predicted that the high resistance state to low resistance state transition was due to the adsorption of gold atom onto a sulfur vacancy leading to the formation of a conducting semi-filament using Nudged Elastic Band (NEB) calculations. However, the dynamics of the electric-field induced filament formation were not considered, which is crucial for predicting realistic shape of filament.

To study the electric-field induced state transition, Mitra et al. [5] performed Reactive Molecular Dynamics simulations for Au/$MoS_2$/Au structures. However, instead of using explicit gold electrodes in their potential, 12/6 Lennard Jones (LJ) walls were used to mimic the electrodes. This approach can result in an artifact due to the absence of realistic interaction of Au and Mo/S atoms. In the subsequent studies [6], grain boundaries in $MoS_2$ were considered to study the filament formation, but the study did not have any experimental validation, hence making the premise of grain boundaries in the initial structure questionable. A recent study [7] uses a deep-learned force-field to carry out molecular dynamics in Au/$MoS_2$/Au memristor, however, lacks electron transport calculations and any experimental validation**.**

In this work, we combine reactive molecular dynamics simulations with *ab-initio* quantum transport and first principles calculations (DFT) to uncover the high-resistance (HRS) and low-resistance states (LRS) in Au-$MoS_2$-Au memristors. We incorporate explicit gold electrodes in the reactive molecular dynamics simulations to study the filament dynamics at the Au/$MoS_2$ interface. We validate our MD insights by using electron transport simulations to compare the computed current densities with reported electrical data.

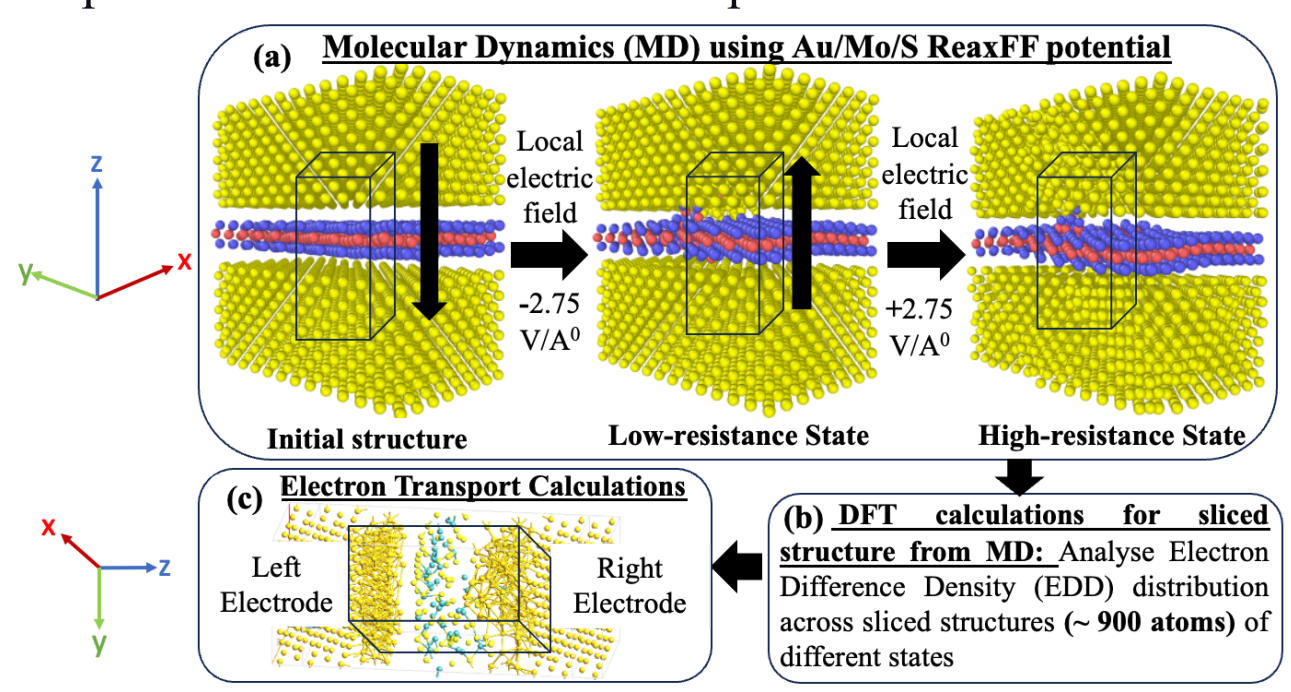

**Fig.1 Simulation methodology combining MD, DFT and quantum transport calculations to model the memristive behavior of Au-$MoS_2$-Au devices**, a) Molecular Dynamics (MD) with ReaxFF potential to obtain electric-field induced states, b) Density-Functional-Theory (DFT) with LCAO basis set to obtain electron density distribution, c) Quantum Transport (QT) within NEGF framework of QuantumATK to obtain the electron current for all three states.

## II. Computational Framework

In this study, the atomic structure of Au-$MoS_2$-Au memristors consists of a 2H phase $MoS_2$ monolayer sandwiched between Au (111) contacts, as shown in Fig. 1a. Molecular dynamics simulations are carried out utilizing LAMMPS [8] to investigate the state transition, as shown in Fig.1 a, b, and c. The interactions between Mo, S and Au atoms are modeled using reactive force field parameters [9], known for its computational efficiency compared to *ab-initio* molecular dynamics, and closely approximating DFT calculations. For MD simulations a timestep of 0.1 fs was used with units set to real and atomic style set to charge. The initial Au-$MoS_2$-Au structure (Fig.2a) is brought in equilibrium under periodic boundary conditions along the x, y, and z axes, employing the NPT ensemble to attain the initial device structure.

Subsequently, the periodic boundary conditions along the x and y axes are altered, and a non-periodic boundary condition along the z axis is introduced to apply electric field along the z-direction. Ambient pressures are set using a Nose-Hoover barostat, while temperature is gradually increased from 10K to 300K using a Berendsen thermostat. Additionally, a sulfur vacancy is introduced to mimic the defect assisted state transitions as reported in experiments [10,11]. The local electric field is applied around the vacancy region to investigate the HRS and LRS structures, as highlighted by the box in Fig. 1(a-c). It is important to note that the value of applied electric fields in MD simulations are much higher than the experiments. This approach mitigates the limitations of MD to capture the resistive switching at larger time scales (> 1 ns) while still capturing the switching kinetics in a computationally effective manner. To understand the relation between applied electric field and switching time, we calculate the switching times (30 ps, 100 ps, 250 ps, 325 ps, 500 ps, 1 ns) for fields (2.75 V/A$^0$, 2 V/A$^0$, 1.5 V/A$^0$, 1.42 V/A$^0$, 1.35 V/A$^0$, 1.3 V/A$^0$). Extrapolating to the experimentally observed switching electric field (~ 0.078V/A$^o$) and voltage (~ 1V) indicates a switching time of 13.22ns aligning with the reported values [10]**.** This method of employing higher state variable values is commonly utilized in MD to reduce simulation timescales while minimizing computational overhead [13,14].

The MD generated structures approximately consist of 5000 atoms, which results in a significantly large computational time for subsequent DFT and ballistic electron Quantum Transport (QT) calculations. Moreover, to validate with the experiments of single-defect $MoS_2$ memristor [10], one of the electrode areas should represent the diameter of the STM tip. Therefore, the active regions of the structures (cross-sectional area of 3.2x5.7 nm$^2$) are identified around the proximity of sulfur defects in $MoS_2$. These sliced structures are used for DFT calculations (~ 900 atoms) and QT calculations (~ 600 atoms) within the NEGF framework in QuantumATK package [12] to compute the electron difference density (EDD) and current densities, respectively. For the exchange correlation functional, the generalized gradient approximation (GGA) with the PBE basis set is employed, supplemented with DFT-D3 correction to account for dispersion interactions. The calculations are performed with a k-point density of 4x4x1, utilizing a density mesh cut-off of 60 Hartree. For NEGF calculations the grid density was set to 2x2x226.

## III. Results and Discussions

We begin the study by performing reactive molecular dynamics simulations. Fig.2 (a,d) depicts the initial relaxed Au/$MoS_2$/Au structure with single sulfur vacancy in the monolayer $MoS_2$ layer. To study the effect of a single defect assisted resistive switching mechanism in $MoS_2$ based memristive devices [11], we apply a local electric field in the vicinity of the sulfur vacancy, as shown in Fig.1(a). To mimic the voltage pulse rise time applied in experiments to cause a state transition, we slowly increased the magnitude of the electric field from -0.5 V/A$^o$ to -2.75 V/A$^o$. It allows some of the gold atoms of the top electrode to gain sufficient kinetic energy and migrate through the Van der Waals gap when the electric field reaches -2.75 V/A$^o$. The movement of gold atoms towards the monolayer of $MoS_2$ results in the formation of a semi-filament, as shown in Fig.2 (b,e). We term this state as a low-resistance state (LRS).This structure shows more realistic representation of the filament as compared to reported studies where a few Au atoms are adsorbed or substituted in place of sulfur vacancies to result in a low resistance state [4,15]. For pristine $MoS_2$ case when an electric field was applied in a similar manner, no gold filament formation was observed.

One interesting observation from this case is that we couldn't observe the formation of a full filament where the gold atoms of the top electrode migrate through the $MoS_2$ layer and connect to the bottom electrode of gold. The reason behind the inability of $MoS_2$ based memristors to form a full filament of gold atoms was the high activation energy of 6.991 eV [4] which is practically not feasible. However the formation of a semi-filament is feasible because of low-barrier energy of 0.378 eV [4]. To understand the bipolar resistive switching mechanism of the device, we apply an electric field along the positive z-axis which is opposite to the direction of electric field applied which had led to formation of the semi-filament. We again increase the magnitude of the electric field slowly from 0.5 V/A$^o$ to 1.75 V/A$^o$. Once the magnitude of the electric field reaches 1.75 V/A$^o$, the gold atoms migrate back towards the top electrode as shown in Fig. 2(c,f). This state is termed the high-resistance state (HRS).

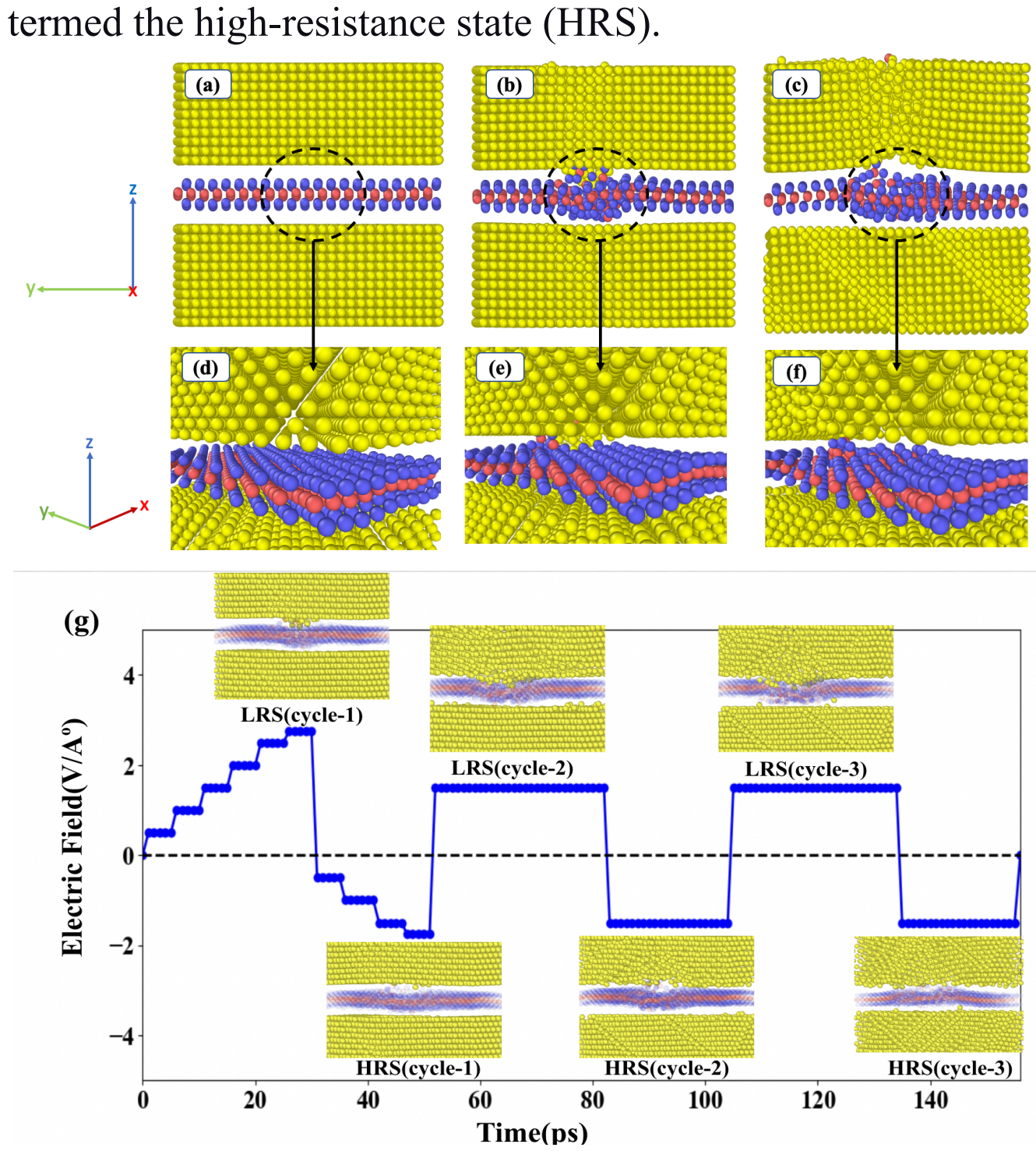

**Fig.2 Electric-field Induced State Transitions in Au-monolayer $MoS_2$-Au system through MD**, 1) From initial state (a,d) to LRS state (b,e) after applying electric field of -2.75 V/A$^o$, where the inset (c,f) shows the semi-conductive filament of gold atoms on $MoS_2$, 2) From LRS to HRS after applying a reverse electric field of 1.75 V/A$^o$, where the inset shows removal of Au atoms from the filament, (g) Variation of electric field with time and the corresponding LRS and HRS structures of cycle 1, 2 and 3.

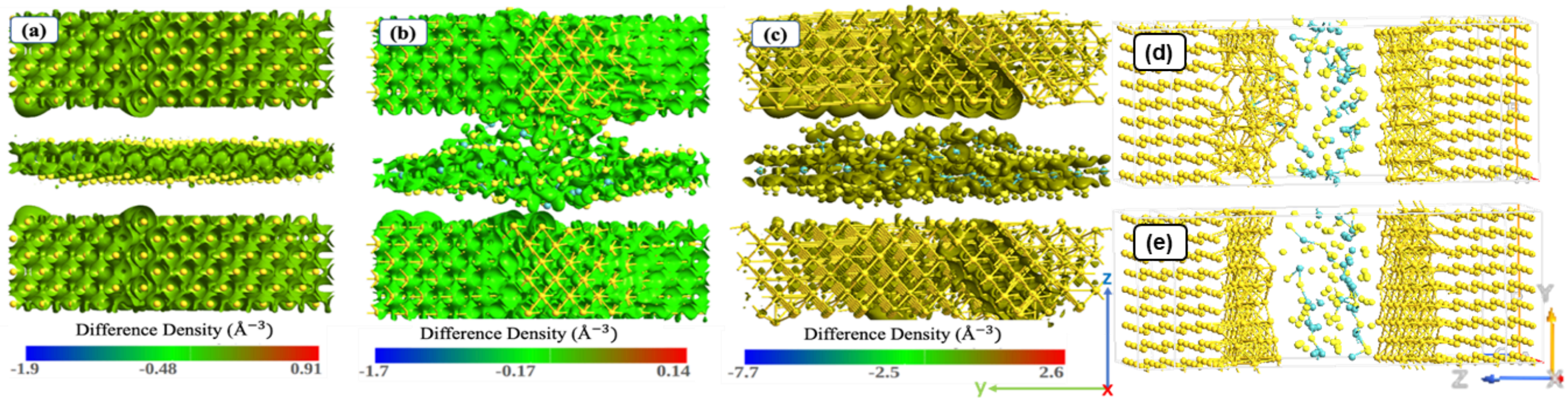

**Fig.3 First-principles calculations,** (a),(b) and (c) show the electron difference density distribution map from DFT of the initial structure, LRS (cycle 1) and HRS (cycle 1) respectively, (d),(e) presents the LRS (cycle 1) and HRS (cycle 1) structures used for *ab-initio* quantum transport, respectively.

Notably, the HRS structure differs significantly from the initial structure shown in Fig. 2(a,d). As shown in Fig. 2(g), the LRS and HRS structures for cycles 2 and 3 were obtained by applying electric fields of -1.5 V/Å and 1.5 V/Å, respectively. Notably, the required field for the HRS-to-LRS transition in cycle 2 is smaller than that for the initial-to-LRS transition in cycle 1.

To understand the electronic difference density distribution (EDD) across the different states we perform DFT calculations. As mentioned earlier in section II, we use smaller structures (~ 900 atoms) for DFT calculations. When we map EDD onto the initial structure (Fig. 3(a)), we observe that there are no charge carriers within the Van der Waals gap indicating a high-resistance path for electrons between the top and bottom electrodes. When we map EDD onto the obtained LRS structure (Fig.3(b)), we see a formation of a conducting channel of electrons between the top electrode and $MoS_2$. This shows that due to the formation of a semi-filament of gold atoms on top of $MoS_2$ a low-resistance path for electrons is created to travel from the top gold electrode to $MoS_2$. Fig.3(c) exhibits the broken electron channel (at the interface of top electrode and $MoS_2$) in the obtained HRS.

In order to validate our single-defect memristor structures with experiments [11], we perform *ab-initio* quantum transport calculations where transmission spectra and current-voltage characteristics are obtained, as shown in Fig. 4. Fig. 4(a) presents the transmission spectrum of LRS which is approximately 1.68 times higher than the HRS when averaged over an energy range of -0.5 to 0.5 eV. Fig. 4(b) shows the comparison of low-bias current densities from quantum transport calculations with single-defect $MoS_2$ memristor data [11] for a lateral STM tip resolution of ~ 0.1 nm. The computed current densities of $8.29 \times 10^7$ A/cm² for LRS and $5.20 \times 10^7$ A/cm² for HRS at 0.1 V are in good agreement with the experimentally reported values (~ $10^7$ A/cm²). The computed conductance values at 0.1 V bias are 137 µS for LRS, 82.5 µS for HRS, and 2.17 µS for the initial structure. This suggests that the initial $MoS_2$ structure with the single defect does not represent the HRS, though it improves the ON/OFF conductance ratio. We observe that both computed current densities and conductances (slope of I-V curves) are higher than the experimental values. It is likely due to the absence of carrier scattering in our ab-initio quantum transport calculations, which has been shown to significantly reduce both current densities and conductances [16]. However, accounting for this is beyond the scope of this study due to the computational challenges involved in incorporating carrier scattering.

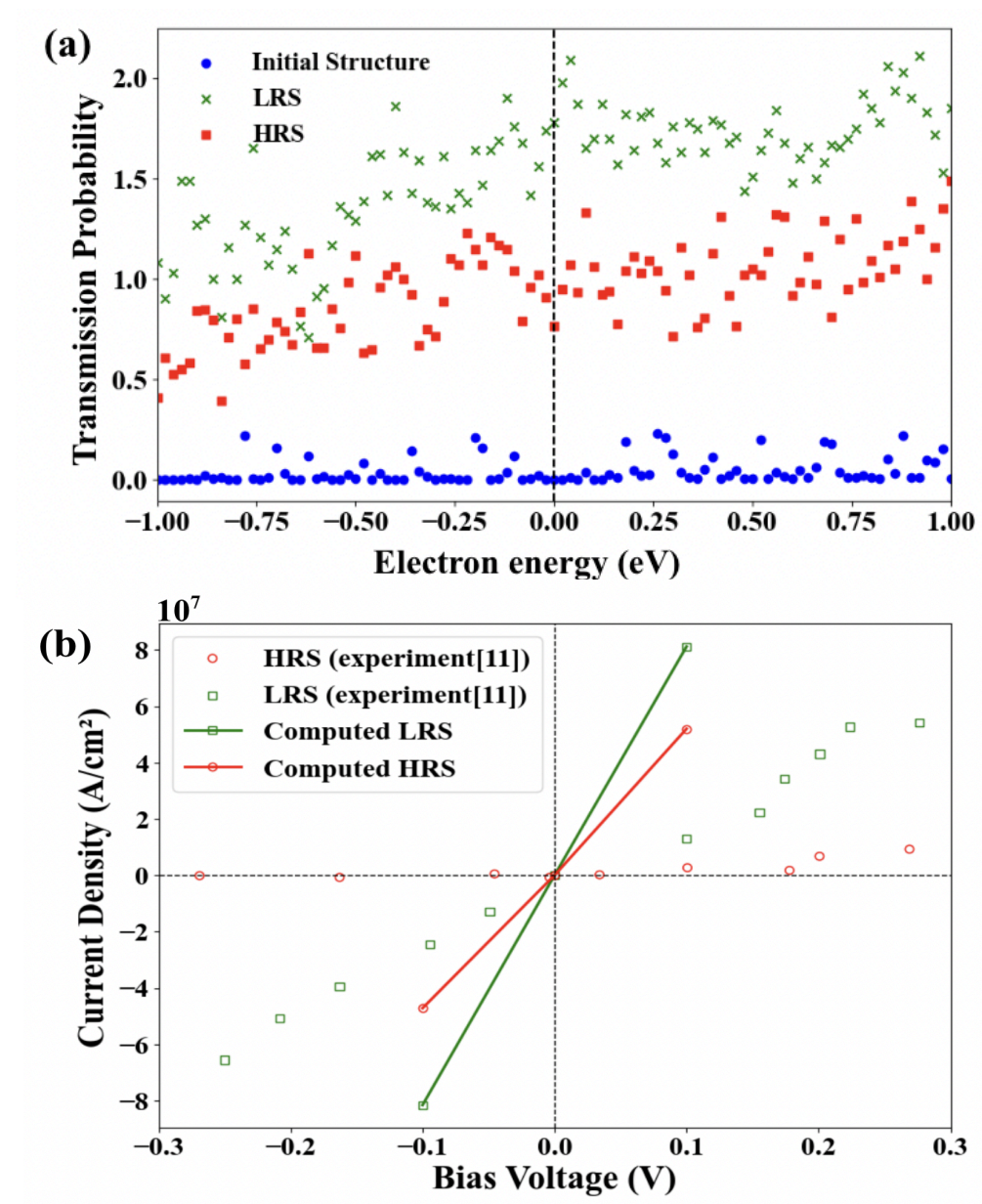

Fig. 4, (a) Transmission spectra of the initial structure, LRS (cycle 1) and HRS (cycle 1), (b) Comparison of current densities with applied bias from simulations & experiments [11].

## IV. Conclusions

In this work, we investigate the resistive switching mechanism in Au/$MoS_2$/Au memristive devices using reactive molecular dynamics, DFT, and electron transport simulations. Reactive molecular dynamics reveal realistic LRS and HRS structures, capturing filament formation and rupture over multiple cycles. These insights into filament dynamics are experimentally validated through electron transport simulations, which confirm the computed current densities against reported experimental data. We demonstrate that the defective $MoS_2$ initial structure does not represent the true HRS, resulting in an HRS-to-LRS resistance ratio below 10, consistent with experimental values for single-defect memristors.